# WKB-approach for the 1D hydrogen atom

A.M. Ishkhanyan[1,2], and V.P. Krainov[3]

[1] Russian-Armenian University, 0051, Yerevan, Armenia

[2] Tomsk polytechnic university 634050, Tomsk, Russia

[3] Moscow Institute of Physics and Technology, 141700 Dolgoprudny, Moscow Region, Russia

e-mail: vpkrainov@gmail.com

Abstract. Taking into account results of WKB-approximation, we derive exact quantum energies and wave functions of even and odd states in the one-dimensional Coulomb potential $V(x) = -1/|x|$, $-\infty < x < \infty$.

1. Introduction

One-dimensional systems with the Coulomb interaction are attractive due to large amount of physical applications. First of all, this is a hydrogen atom in the superstrong magnetic field [1-3]. The other application is the determination of the electron bound energies near the surface of liquid $He^3$, Ne and solid molecular deuterium [4-5]. States of donors in the single-wall carbon nanotubes were considered in Ref. [6].

2. Theory

The stationary Schrödinger equation is of the form (in atomic system of units):

$$-\frac{1}{2}\frac{d^2\psi_n}{dx^2} - \frac{\psi_n}{|x|} = E_n\psi_n; \quad E_n < 0; \quad -\infty < x < \infty. \quad (1)$$

This is so called Whittaker equation. Its solutions for bound even and odd states are given by the Whittaker functions which are zero in the infinity:

$$\psi_n(x) = \left(signum(x)\right)^n W_{(n+1)/2,1/2}\left(\frac{4|x|}{n+1}\right); \quad E_n = -\frac{2}{(n+1)^2}. \quad (2)$$

The function (2) is symmetrical for even values of $n$ and anti-symmetrical for odd $n$.

The difficulties exist for even solutions. The wave function of these solutions are nonzero in the origin, so that it is impossible to determine the single-valued quantum energies. If we exclude the singularity in the origin of the Coulomb potential, for example, as $|x| \to |x| + a$, $a \ll 1$, then the energy of the ground state is approximately [7] $E_0 \approx -2\ln^2\frac{1}{a}$.

This quantity diverges when $a \to 0$. If the Coulomb potential with the repulsive core is considered [8]: $V(x) = -\frac{|x|-b}{(|x|+a)^2}$, $a,b \ll 1$, then at $1 < \frac{b}{a} < \ln\frac{1}{a}$ even states are found between odd states, as should be according to the principles of quantum mechanics. In Ref. [9] bound electron states near the surface of liquid helium are described using the simple potential $V(x) = -\frac{1}{x}; x > 0; \quad V(x) = \infty; x < 0$. But in this case odd well-known solution of Eq. (2) remain . The additional potential which is proportional to the Dirac delta-fuction $\delta(x)$, was introduced in Ref. [10], in order to find even solutions. The energy spectrum of the relativistic Dirac particle in the 1D Coulomb potential $V(x) = -1/(|x|+a)$, $a \ll 1$, was considered in Ref. [11]. Some other types of Hamiltonian modification near the origin were suggested in [12-14].

Here we would like first of all to underline, that when we use the WKB-approximation, i.e. the Bohr quantization rule, there are no infinities for energies of atomic states. Indeed, the WKB-energies are given by the well-known Balmer formula:

$$\int_{-1/|E_n|}^{1/|E_n|} \sqrt{2\left(-|E_n| + \frac{1}{|x|}\right)} dx = (n+1)\pi; \quad n = 0, 1, 2, 3...; \quad E_n = -\frac{2}{(n+1)^2}. \tag{3}$$

The Maslov index is equal to one in this quantization rule due to the Coulomb singularity in the origin [15]. For highly-excited states with $n \gg 1$ the exact values of energies in (2) should be coincide with the semiclassical values (3). This can be seen from their comparison both of even and odd states. We can assume that the same will be for moderate values of quantum numbers. It is seen from (2) and (3), that they coincide with each for all even and odd states analogously to the same situation for 3D hydrogen atom.

For the ground (even) state it follows from comparison of (2) with (3) that $E_0 = -2; \quad n = 0$. The wave function (non-normalized) of the ground state $W_{1/2,1/2}(4|x|)$ is shown in Fig. 1. It does not contain nozzles, as it should be. Its value in the origin is finite and equal to 0.5642… The infinite derivative in the origin (cusp) reflects logarithmic singularity.

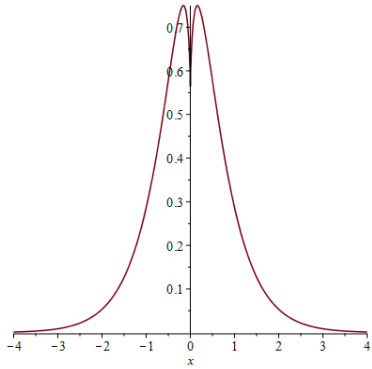

Fig.1. The wave function of the ground 1D hydrogen state ($n = 0$).

The first excited (odd) state has the energy $E_1 = -1/2$; $n = 1$. Its wave function contains one nozzle (at $x = 0$) and is depicted in Fig 2. Obviously, this wave function coincides with the wave function of the ground state of the 3D hydrogen atom. Even and odd states alternate with each other.

The second excited (even) state has the energy $E_2 = -2/9$; $n = 2$. Its wave function is shown in Fig. 3 and contains two nozzles. All even wave functions contain cusp in the origin, unlike the odd states. Of course, wave functions and energies of odd states coincide with wave functions and energies of the 3D Coulomb problem with zero orbital momentum.

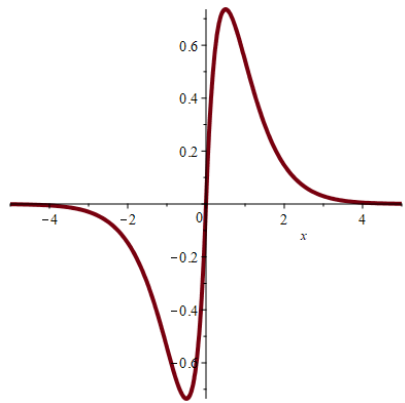

Fig.2. The wave function of the first excited state of the 1D hydrogen atom ($n = 1$)
$$\psi_1 \propto x \exp(-|x|)$$

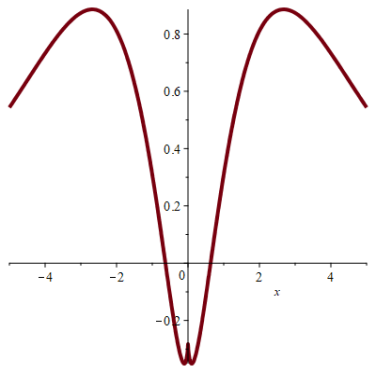

Fig.3. The wave function of the second excited 1D hydrogen state ($n = 2$).

3. Conclusion

Thus, we have found energies and wave functions of all even and odd states of the one-dimensional hydrogen atom without introduction of additional potentials near the Coulomb singularity.

The work has been supported by the Armenian State Committee of Science (SCS Grant No. 18RF-139), the Armenian National Science and Education Fund (ANSEF Grant No. PS-4986), the Russian-Armenian (Slavonic) University at the expense of the Ministry of Education and Science of Russian Federation, the project 'Leading Russian Research Universities' (Grant No. FTI_24_2016 of the Tomsk Polytechnic University), the Russian Foundation for Basic Research (project № 18-52-05006), and the Ministry of Education and Science of Russian Federation (project № 3.873.2017/4.6).

666